\documentclass[prstab,showpacs,preprintnumbers,amsmath,amssymb,showkeys]{revtex4}

\usepackage{graphicx}% Include figure files
\usepackage{bm}% bold math
\begin{document}
\title{Nuclear medium effects in $\nu/\bar{\nu}$-A DIS}

%\keywords{Nuclear medium effects, Neutrino-nucleus interactions, Structure functions, Deep inelastic scattering, Local density approximation}
\author{H. Haider$^*$}
\author{I.Ruiz Simo$^!$}
\author{M.Sajjad Athar$^*$}
\author{M. J. Vicente Vacas$^!$}
\affiliation{$^*$Department of Physics, Aligarh Muslim University, Aligarh-202 002, India}
\affiliation{$^!$Departamento de F\'{\i}sica Te\'orica and IFIC, \\
Centro Mixto Universidad de
Valencia-CSIC,\\
46100 Burjassot (Valencia), Spain}

\begin{abstract}
 Nuclear medium effects in the weak structure functions $F_2(x,Q^2)$ and $F_3(x,Q^2)$ have been studied for deep inelastic neutrino/antineutrino reactions in iron nucleus by taking 
into account Fermi motion, binding, pion and rho meson cloud contributions, target mass correction, shadowing and anti-shadowing corrections. The calculations have been performed 
in a local density approximation
 using relativistic nuclear spectral functions which include nucleon correlations. Using these structure functions we have obtained the ratio
 $R_{F2,F3}^A(x,Q^2)=\frac{2F_{2,3}^A(x,Q^2)}{AF_{2,3}^D(x,Q^2)}$, the differential scattering cross section
 $\frac{1}{E}\frac{d^2\sigma}{dxdy}$ and the total scattering cross section $\sigma$.
The results of our numerical calculations in $^{56}Fe$ are compared with the experimental results of NuTeV and CDHSW collaborations. 
\end{abstract}
\maketitle
\section{Introduction}
Presently many experiments are going on and some are under way to study 
neutrino events using atmospheric as well as accelerator neutrinos in the energy region of few GeV. 
Several new experiments using various neutrino sources like beta beams,
 super-beams or neutrinos from factories have been discussed in literature. 
In most of these experiments nuclear targets are being used. In this energy range, contribution to the
$\nu$($\bar\nu$)-nucleus cross section comes from quasi-elastic, inelastic as well as deep inelastic processes.
\\
For the deep inelastic neutrino-nucleus scattering there are only a few calculations 
where the dynamical origin of the nuclear medium effect 
has been studied, and in some theoretical analysis, nuclear medium effect has been 
phenomenologically  described in terms of a few parameters 
which are determined by fitting the experimental data of charged leptons and 
neutrino deep inelastic scattering(DIS) from various nuclear targets.
\\
In this paper, we have studied ratio of the structure functions $F_2^A$ and $F_3^A$ in iron and deuterium nuclei and the differential as well as total scattering 
cross section in iron. These results have been obtained in a
 theoretical model~\cite{oset611,athar,athar1} using spectral function~\cite{oset1697} to describe the momentum distribution 
of nucleons in the nucleus. The spectral function has been
 calculated using the Lehmann's representation for the relativistic nucleon propagator, nuclear many body theory is then applied to calculate
 it for an interacting Fermi sea in nuclear matter~\cite{oset611}.
 A local density approximation is then 
applied to translate these results to finite nuclei.
We have taken the effect of pion and rho meson contributions~\cite{oset611}, target mass correction~\cite{schienbein}, 
and nuclear shadowing and anti-shadowing effects~\cite{kulagin}. We have used CTEQ6.1~\cite{cteq} for parton distribution functions(PDFs).
The QCD evolution of the structure functions have also been taken into account using the coefficient functions given
 in Refs.~\cite{vermaseren, Moch}. Numerical results are compared with the experimental results of NuTeV~\cite{nutev} and CDHSW~\cite{cdhsw} collaborations, 
and also these numerical results are compared with the results obtained from the various phenomenological studies~\cite{eskola,hirai,Schienbein1}. 
% &&&&&&&&&&&&&&&&&&&&&&&&&& REsults &&&&&&&&&&&&&&&&&&&&&&&&&&&&&&&&&&&&&&&&&&&&&&&&&&&&&&&&&&&&&&&&&&&7
\section{formalism}
The expression of the differential cross section, for deep inelastic scattering (DIS) of neutrino with a nucleon target induced by charged current reaction 
\begin{equation} 	\label{reaction}
\nu_l(k) + N(p) \rightarrow l^-(k^\prime) + X(p^\prime),~l=~e,~\mu
\end{equation}
is given by
\begin{footnotesize}
\begin{eqnarray} \label{cross_section}
\frac{d^2\sigma^{\nu(\bar{\nu})}}{dx\ dy} &=& \frac{G_F^2 M
E_{\nu}}{\pi(1+Q^2/M_W^2)^2}\Biggl((y^2 x + \frac{m_l^2 y}{2 E_{\nu} M})
F_1(x,Q^2) +\left[ (1-\frac{m_l^2}{4 E_{\nu}^2})
-(1+\frac{M x}{2 E_{\nu}}) y\right]F_2(x,Q^2) \\ \nonumber
&
\pm&
\left[x y (1-\frac{y}{2})\right.
-\left.\frac{m_l^2 y}{4 E_{\nu} M}\right]
F_3(x,Q^2) 
+ \frac{m_l^2(m_l^2+Q^2)}{4 E_{\nu}^2 M^2 x} F_4(x,Q^2)
- \frac{m_l^2}{E_{\nu} M} F_5(x,Q^2)\Biggr)\
\end{eqnarray}
\end{footnotesize}
where $G_F$ is the Fermi coupling constant, M is the mass of nucleon, in $F_3$, +sign(-sign) is for neutrino(antineutrino), x($=\frac{Q^2}{2M\nu}$) is the Bjorken variable, $y=\frac{\nu}{E}$, $\nu$ and q being the energy
and momentum transfer of leptons and $Q^2=-q^2$. $F_4$ and $F_5$ are generally omitted since they are suppressed by a factor of at least
$m_l^2/2ME_\nu$ relative to the contributions of $F_1$, $F_2$ and
$F_3$. $F_1$ and $F_2$ are related by the Callan-Gross relation~\cite{Callan} leading to only
two independent structure functions  $F_2$ and $F_3$.

The expressions of $F_2$ and $F_3$ structure functions due to nuclear medium effect like Fermi motion and binding
energy in the present model are given by~\cite{oset611,athar,athar1}
\begin{eqnarray}\label{f2Anuclei}
F^A_2(x_A,Q^2)&=&4\int d^3r\int \frac{d^3p}{(2\pi)^3}\frac{M}{E({\bf p})}\int_{-\infty}^\mu dp^0\;
S_h(p^0,\mathbf{p},\rho(\mathbf{r})) \frac{x}{x_N}
\left( 1+\frac{2x_N p_x^2}{M\nu} \right)  F_2^N(x_N,Q^2)~~~~~
\end{eqnarray}
\begin{eqnarray}\label{f3Anuclei}
F_3^A(x_A,Q^2)&=&4\int d^3r \int \frac{d^3p}{(2\pi)^3} \frac{M}{E({\bf p})}\int_{-\infty}^{\mu} dp^0
S_h(p^0,\mathbf{p},\rho(\mathbf{r})) \frac{p^0\gamma-p_z}{(p^0-p_z\gamma)\gamma} F_3^N(x_N,Q^2)~~~~~
\end{eqnarray}
where 
\begin{equation}	\label{gamma}
\gamma=\frac{q_z}{q^0}=
\left(1+\frac{4M^2x^2}{Q^2}\right)^{1/2}\,,
\end{equation}
and \[x_N=\frac{Q^2}{2(p_0q_0-p_zq_z)}\]
$S_h(p^0,\mathbf{p},\rho(\mathbf{r}))$ is the nuclear spectral function the expression for which is taken 
from ref.~\cite{oset1697}. We ensure that the spectral function is properly normalized and we get the 
correct Baryon number and binding energy for the nucleus.

Above expressions of structure functions are our base equations. To our base equations we have incorporated other medium effects namely  pion and rho cloud contributions following Ref.~\cite{oset611},
 and shadowing and anti-shadowing effects using the prescription given in Ref.~\cite{kulagin}, some details of which
are given in Ref.~\cite{athar1} . Using Eqns.\ref{f2Anuclei} and \ref{f3Anuclei} in Eq.\ref{cross_section}, we have calculated the total scattering cross section.
\section{Results and Discussion}
\begin{figure}
\includegraphics[height=.3\textheight,width=0.8\textwidth]{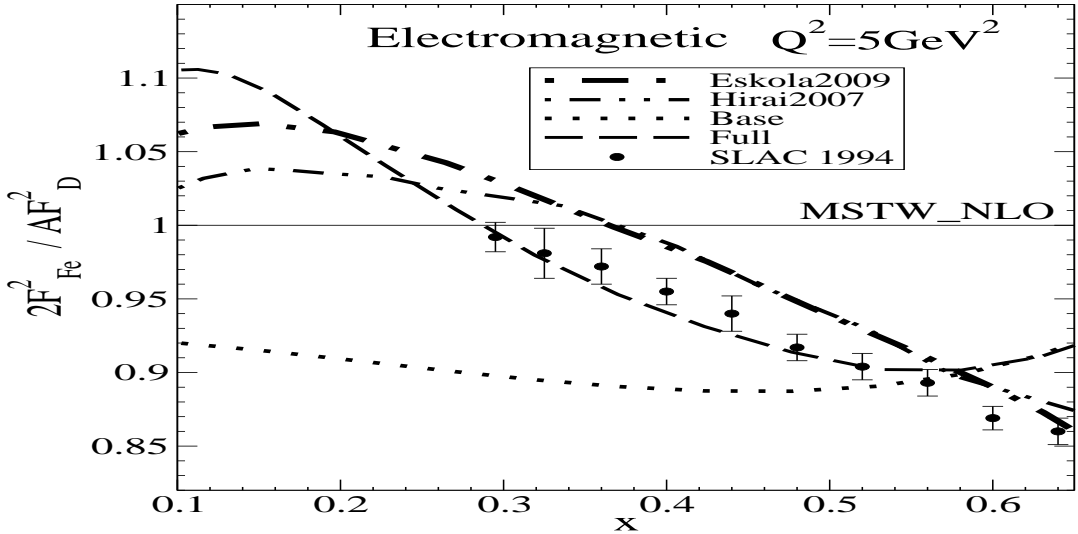}
\caption{$2F_2^{Fe}/AF_2^D$ as a function of x. Present model is compared to SLAC data~\cite{SLAC}. Ratio from base and full calculations are shown by dotted and dashed curves. 
The dotted-dashed and double
dotted-dashed curves represent the phenomenological studies of Eskola et al.~\cite{eskola} and Hirai et al.~\cite{hirai}.}
\label{fig:f2em}
\end{figure}

\begin{figure}
\includegraphics[height=.3\textheight,width=0.8\textwidth]{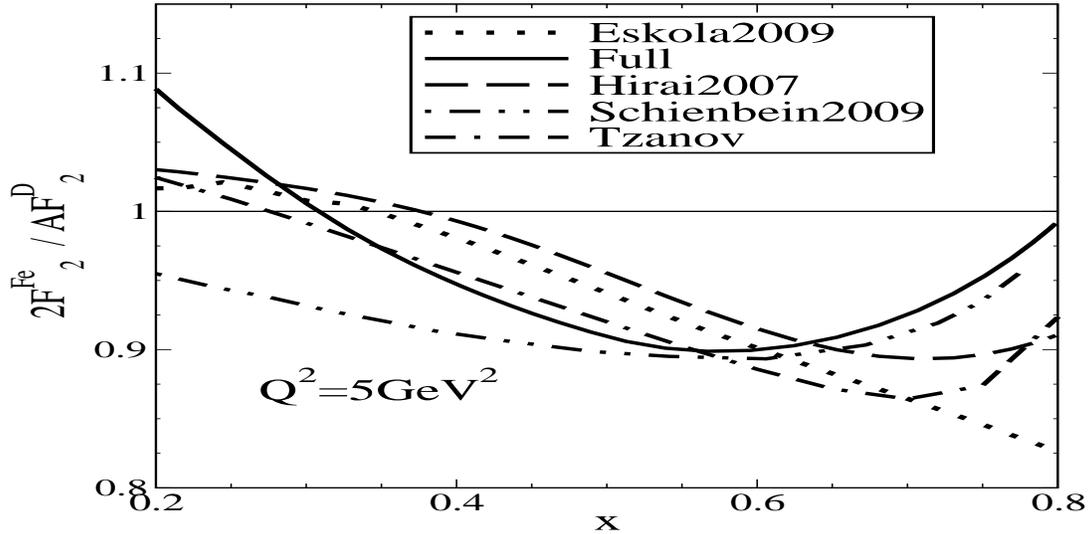}
\caption{Ratio R(x,$Q^2$)=$\frac{2F_{2}^{Fe}}{AF_2^D}$ with full calculation has been shown by the solid line.  
 The results obtained from the phenomenological studies of Tzanov et al.~\cite{nutev}(double dashed-dotted line),
Eskola et al.~\cite{eskola}(dotted line), 
Hirai et al.~\cite{hirai}(dashed line) and Schienbein et al.~\cite{Schienbein1}(double dotted-dashed line) have also been shown.}
\label{fig:f2weak}
\end{figure}

\begin{figure}
\includegraphics[height=.3\textheight,width=0.8\textwidth]{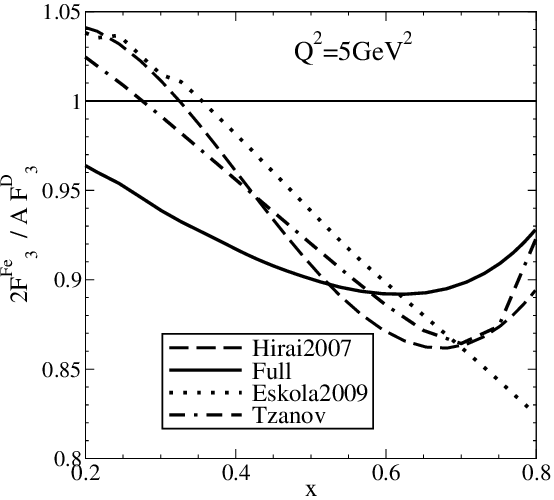}
\caption{Ratio R(x,$Q^2$)=$\frac{2F_{3}^{Fe}}{AF_3^D}$ with full calculation has been shown by the solid line.  
 The results obtained from the phenomenological studies of Tzanov et al.~\cite{nutev}(double dashed-dotted line), 
Eskola et al.~\cite{eskola}(dotted line) and Hirai et al.~\cite{hirai}(dashed line) have also been shown.}
\label{fig:f3weak}
\end{figure}

\begin{figure}
\includegraphics[height=.3\textheight,width=0.9\textwidth]{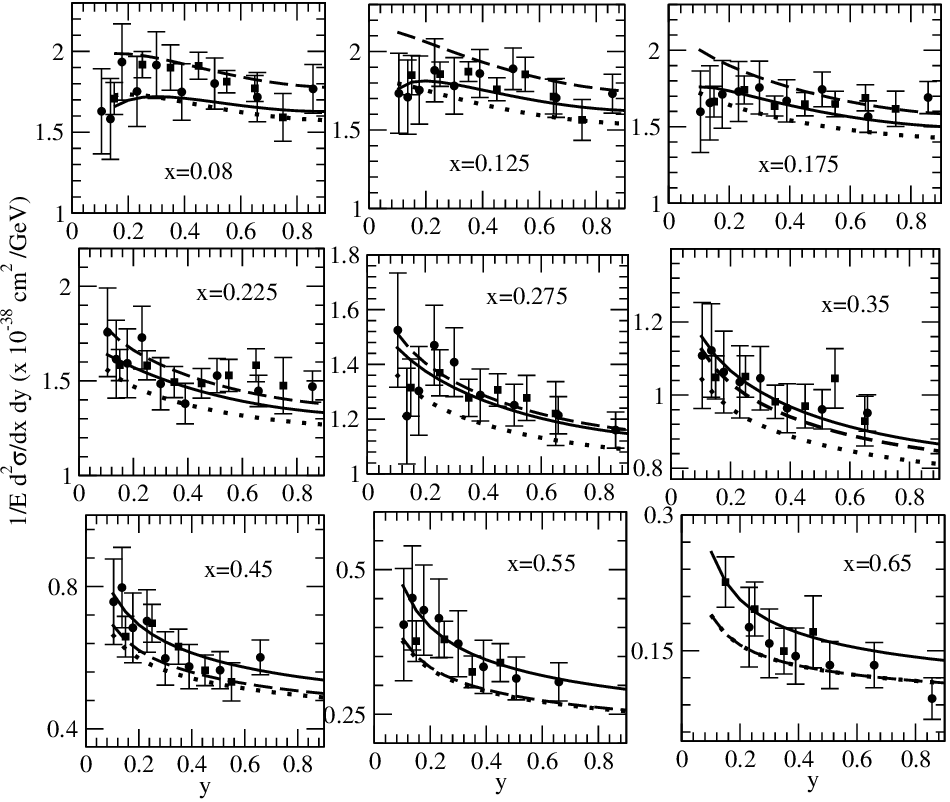}
\caption{$\frac{1}{E}\frac{d^2\sigma}{dxdy}$ vs y at different x for $\nu_\mu$($E_{\nu_\mu}=65GeV$) induced reaction in $^{56}$Fe.
 Dotted(Dashed) line is the result with base(full) calculation at LO. Solid line 
is the result with the full calculation at NLO. NuTeV~\cite{nutev} results have been shown by the solid squares and CDHSW~\cite{cdhsw} results by solid circles.}
\label{fig:d2sig_neut_65gev}
\end{figure}

\begin{figure}
\includegraphics[height=.3\textheight,width=0.9\textwidth]{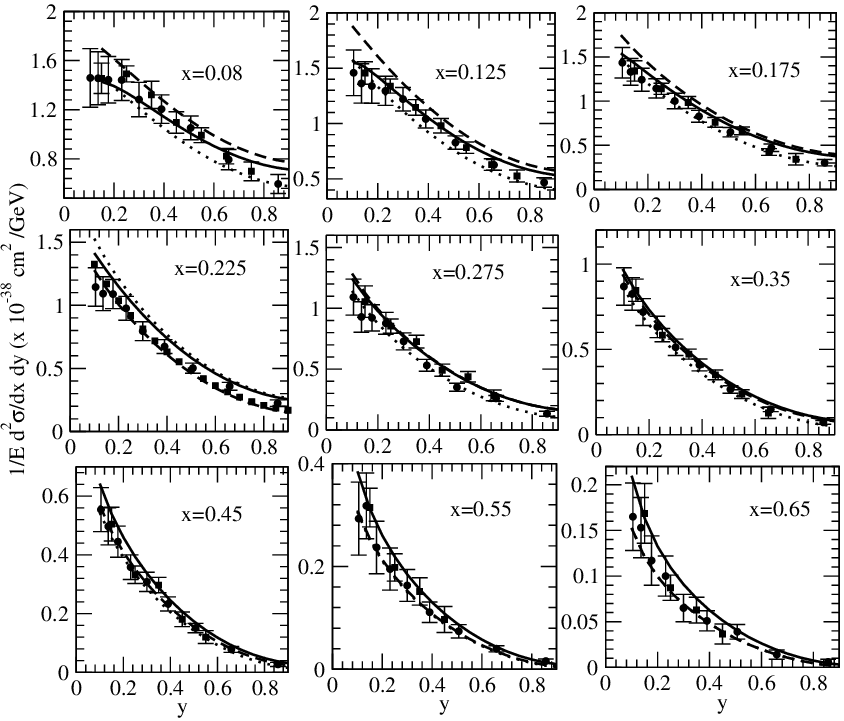}
\caption{$\frac{1}{E}\frac{d^2\sigma}{dxdy}$ vs y at different x for $\bar \nu_\mu$($E_{\bar\nu_\mu}=65GeV$) induced reaction in $^{56}$Fe for antineutrino.
 Dotted(Dashed) line is the result with base(full) calculation at LO. Solid line 
is the result with the full calculation at NLO. NuTeV~\cite{nutev} results have been shown by the solid squares and CDHSW~\cite{cdhsw} results by solid circles.}
\label{fig:d2sig_antineut_65gev}
\end{figure}

\begin{figure}
\includegraphics[height=.3\textheight,width=0.8\textwidth]{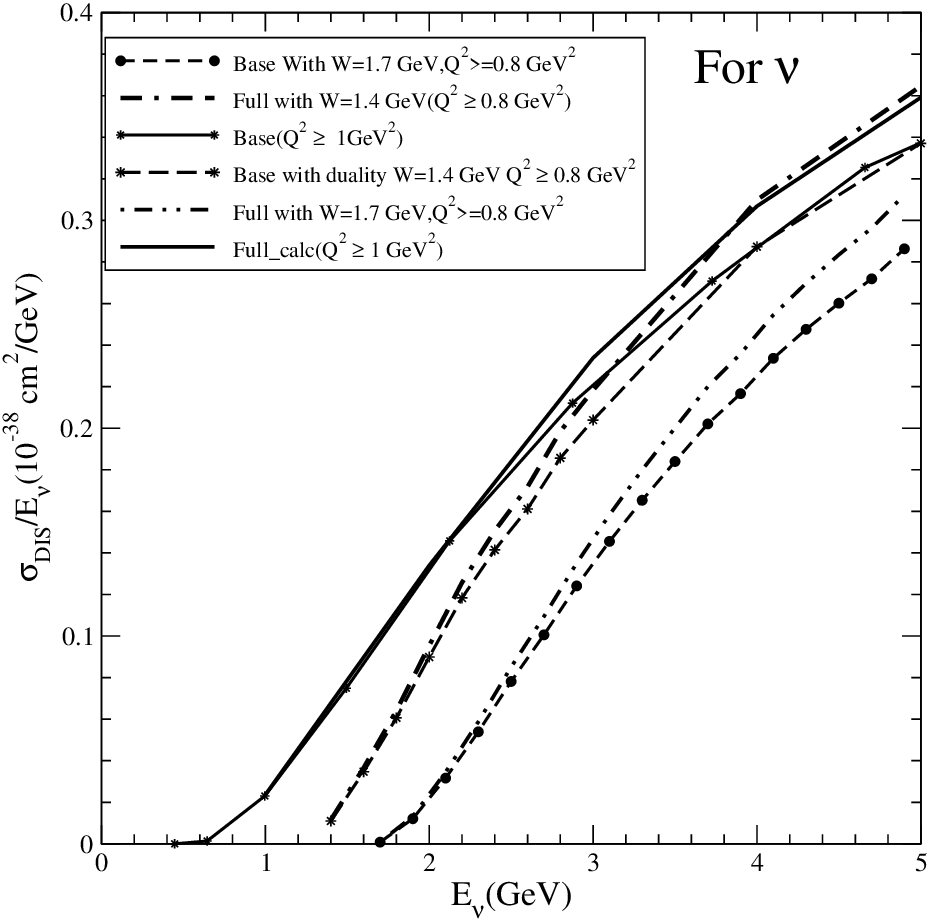}
\caption{$\sigma$ vs $E_\nu$ using iron target for neutrino induced process using CTEQ6.1 PDFs at LO. $\sigma$ is normalized to $1/E_\nu$.
 Solid(solid with stars) line is the result with full(base) calculations for $Q^2\geq 1 GeV^2$ and without duality cut. Dashed dotted and dashed with stars lines
are the results with the full and base calculations respectively for $Q^2\geq 0.8 GeV^2$ and W=1.4 GeV.
Double dotted-dashed line and dashed line with circles are curves plotted for full and base calculations respectively with W=1.7 GeV and $Q^2\geq 0.8 GeV^2$.}
\label{fig:cross_section_neutrino}
\end{figure}

\begin{figure}
\includegraphics[height=.3\textheight,width=0.8\textwidth]{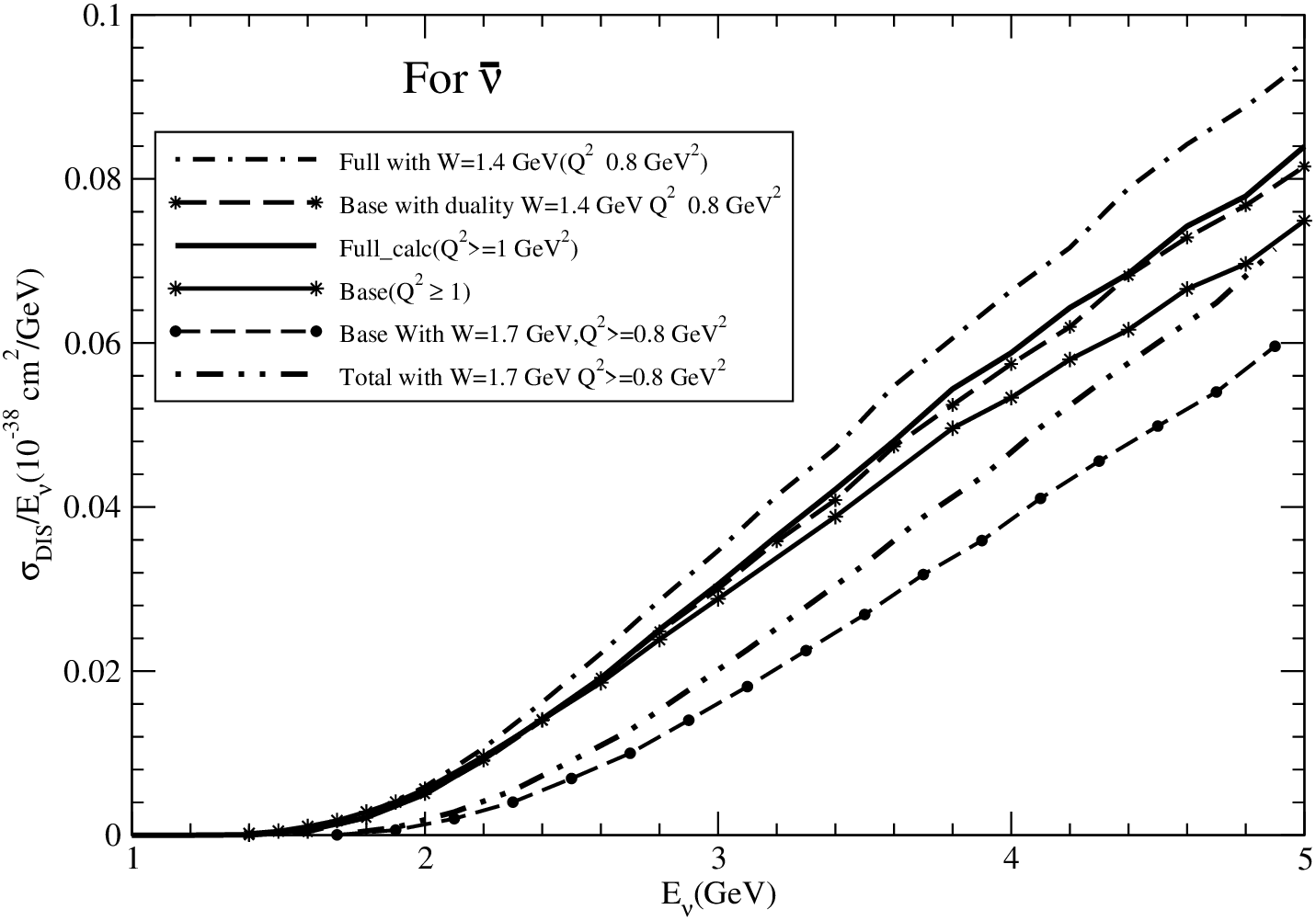}
\caption{$\sigma$ vs $E_\nu$ using iron target for antineutrino induced process.Lines have the same meaning as in fig. \ref{fig:cross_section_neutrino}.}
\label{fig:cross_section_antineutrino}
\end{figure}

Recently, we have studied the effect of nuclear medium on the electromagnetic nuclear structure function $F_2(x,Q^2)$ in nuclei by including nucleonic and mesonic degrees of freedom, 
shadowing and anti-shadowing corrections,
 Fermi motion and binding effects~\cite{athar1}. We have obtained the ratio R(x,Q$^2$)=$\frac{2F_{2}^{A}}{AF_2^D}$ for several nuclei and compared our results with the 
recent JLab results of Ref.~\cite{Seely} as well as from some of the older experiment~\cite{SLAC}.
In fig.\ref{fig:f2em}, we have presented the ratio of structure function in iron and deuteron(Eq.(60) of Ref.~\cite{ciofi}) for electromagnetic interactions and then compared the ratio R with the SLAC data ~\cite{SLAC}.
We have obtained the same ratio in weak case at NLO and depicted in figs. \ref{fig:f2weak} and \ref{fig:f3weak}.
 We find that the $F_2$ ratios in weak and electromagnetic cases are of similar nature while $F_3$ is of different nature. 
Results for the differential scattering cross section for neutrino and antineutrino induced reactions have been presented in fig. \ref{fig:d2sig_neut_65gev} and fig.\ref{fig:d2sig_antineut_65gev}
respectively obtained by using Eq.(\ref{cross_section}) with CTEQ6.1 PDFs and $Q^2>1GeV^2$. On comparing these numerical results with experimental data from NuTeV and CDHSW collaboration,
 we find that our results are fairly in good agreement with experimental data and calculations at NLO make results better. These results are presented using our base equation and with full calculations where pion and rho meson contributions and nuclear 
shadowing and anti-shadowing effects are taken into account. In figs.\ref {fig:cross_section_neutrino} and \ref{fig:cross_section_antineutrino}, we have presented the results for the total scattering cross 
section $\sigma$ for charged current neutrino and antineutrino induced processes with and without center of mass energy(W) cut.  We find that in the neutrino energies of around 2-5GeV when the Fermi motion and binding effects are taken into
 account the cross section reduces from the free case(not shown here) which is around $8-9\%$. When shadowing and antishadowing, pion and rho meson cloud contributions are taken into account which corresponds 
to the full calculation, the cross section increases by about $4-6\%$ from the base results. Therefore, the net effect
of medium correctios is $4-5\%$ reduction from the cross sections calculated for the free case. To show the effect of cut on the center of mass energy which is better known as the 
duality cut on the cross section, we have taken two different values of W (1.4GeV and 1.7GeV) for the same $Q_{min}^2$.
 We find that when the cross sections calculated with the duality cut of 1.4GeV and $Q^2>0.8GeV^2$, the cross sections
 reduces by about $28\%$ at 2GeV and $1\%$ at 4GeV, while if we apply a duality cut of 1.7GeV the cross sections reduces
 by about $50\%$ at 2GeV and $6\%$ at 4GeV as compared to the cross sections calculated without the duality cut and $Q_{min}^{2}>1GeV^2$.
%&&&&&&&&&&&&&&&&&&&&&&&&&&&&&&&&&&&&&&&&&&&&&&&&&&&&&&&&&&&&&&&&&&&&&&&&&&&&&&&&&&&&&&&&&&&&&&&&&&&&&&&&&&&&&&&&&&&&&&&&&&&
%\bibliographystyle{aipproc}   

\end{document}